\documentclass[runningheads,a4paper]{llncs}

\usepackage{amssymb}
\usepackage{graphicx}
\graphicspath{{./Figures/}}   % Location of your graphics files

\usepackage{url}
\usepackage{bsymb}
\usepackage{hyperref}
\hypersetup{colorlinks=true}   % Set to false for black/white printing
\usepackage{subfigure}
\usepackage{eventB}
\usepackage{xcolor}

\urldef{\mailsa}\path|ras07r@ecs.soton.ac.uk|
\newcommand{\keywords}[1]{\par\addvspace\baselineskip
\noindent\keywordname\enspace\ignorespaces#1}

\begin{document}

\mainmatter  % start of an individual contribution

% first the title is needed
\title{Lessons Learned/Sharing the Experience of Developing a Metro System Case Study}

% a short form should be given in case it is too long for the running head
\titlerunning{Sharing the Experience of Developing a Metro System Case Study}

% the name(s) of the author(s) follow(s) next
%
% NB: Chinese authors should write their first names(s) in front of
% their surnames. This ensures that the names appear correctly in
% the running heads and the author index.
%
\author{Renato Silva%
%\thanks{Please note that the LNCS Editorial assumes that all authors have used
%the western naming convention, with given names preceding surnames. This determines
%the structure of the names in the running heads and the author index.}%
%\and Ursula Barth\and Ingrid Haas\and Frank Holzwarth\and\\
%Anna Kramer\and Leonie Kunz\and Christine Rei\ss\and\\
%Nicole Sator\and Erika Siebert-Cole\and Peter Stra\ss er
}

\authorrunning{Sharing the experience of developing a Metro System Case Study}
% (feature abused for this document to repeat the title also on left hand pages)

% the affiliations are given next; don't give your e-mail address
% unless you accept that it will be published
\institute{University of Southampton,\\
%Tiergartenstr. 17, 69121 Heidelberg, Germany\\
\mailsa\\
%\mailsb\\
%\mailsc\\
%\url{http://www.ecs.soton.ac.uk/}
}

%
% NB: a more complex sample for affiliations and the mapping to the
% corresponding authors can be found in the file "llncs.dem"
% (search for the string "\mainmatter" where a contribution starts).
% "llncs.dem" accompanies the document class "llncs.cls".
%

\toctitle{Lessons Learned and Sharing of Experiences While Developing a Metro System Case Study}
\tocauthor{Renato Silva}
\maketitle
\pagestyle{empty}
\begin{abstract}
In this document we share the experiences gained throughout the development of a metro system case study. The model is constructed in Event-B using its respective tool set, the Rodin platform. Starting from requirements, adding more details to the model in a stepwise manner through refinement, we identify some keys points and available plug-ins necessary for modelling large systems (requirement engineering, decomposition, generic instantiation, among others), which ones are lacking plus strengths and weaknesses of the tool.
\keywords{Event-B, Rodin, requirements, refinement, decomposition, generic instantiation}
\end{abstract}
\section{Introduction}
%
%Sharing of the experience while developing a metro system model. Start with building requirements and which plug-in can be use to achieve that. Then focus on abstraction (no easy or tool to help with that), why to use refinement, why to use decomposition, why to use generic instantiation. In the end (if there is time and space), talk about enabledness properties that can be perceived with ProB.

%It is believed that reusability in formal development should reduce the time and cost of formal modelling within a production environment. 
Event-B~\cite{Abrial2010Modeling-in-Eve} is a formal method that allows modelling and refinement of systems. From the experiences during DEPLOY\footnote{DEPLOY - Industrial deployment of system engineering methods providing high dependability and productivity - supported by the EU Commission (Grant 214158)}, there exists a natural instinct to model a system such that it mimics its implementation. That is not always the best approach: models should be used to understand the system and its behaviour; the implementation should be seen as an independent task. %instead modelling and implementation tasks should complement each. 
%Frequently two kind of formal modelling design are used: "top-down'', that starts with an abstract model of the envisaged system and then through refinement, the initial model becomes less abstract and more concrete, closer to an implementation; "bottom-up'', useful for modelling service composition where modules already exist before they are composed. These two design styles try to take advantage of reusability that has always been sought in several areas (e.g. software, mathematics) as a way to reduce time, cost and improve the productivity of developments. %\cite{Standish1984An-Essay-on-Sof}. 
This document aims to guide modellers by describing the experiences gained throughout the development of a metro system case study, suggesting ``rules of thumb'', modelling techniques and assessing the current tool support (Rodin platform~\cite{linkRodin}).

We build a metro system model in a ``top-down'' style, in Event-B based on safety properties, starting from an abstraction view of the system and gradually augmented it with more details. Generic instantiation~\cite{RefDecInst,Silva2009SupportingGenericInst,10.1109/SEFM.2009.17} and decomposition~\cite{Silva2010Decomposition-T} are techniques used in the case study, simplifying the formal development by reusing existing models and avoiding re-proofs.
%We combine shared event decomposition (where sub-components interact via synchronised shared events and shared states are not allow), generic instantiation and refinement to model particular aspects of the system. 
Some requirements are based on real ones for metro system carriage doors.

A brief overview of the Event-B language is given in Section~\ref{Section:Background}. The construction of the metro system model is described in Section~\ref{Section:CaseStudyConstruction}, including a discussion of the keys points for building of a formal model such as requirements, abstraction, refinement, proofs, decomposition, generic instantiation in the Rodin platform. We finish with conclusions and related work in Section~\ref{Section:Conclusions}.

\section{Background}
\label{Section:Background}
Event-B is a formal modelling method for developing \emph{correct-by-construction} hardware and software systems. An Event-B specification is divided into two parts: a static part called \textit{context} and a dynamic part called \textit{machine}. A machine \textit{SEES} as many contexts as desired. A context consists of sets, constants and assumptions (axioms) of the system. 
%Sets in the context can be seen as a collection of elements or a type definition. 
An Event-B model is a state transition system where the state corresponds to \emph{variables} $v$ and transitions are represented by a
collection of \emph{events} $\Bevt{evt}$ in machines. The most general form of an event is: $\inlineevent{evt}{t}{G(t,v)}{S(t,v,v')}$, where $t$ is a set of parameters, $G(t,v)$ is the enabling condition (called guard) and $S(t,v,v')$ is a before-after predicate computing after state $v'$. Essential to Event-B is the formulation of \emph{invariants} $I(v)$: safety conditions/properties to be preserved at all times. \emph{Proof obligations} (PO) are generated for all system transitions to validate and ensure that these conditions are preserved. Because Event-B advocates the use of refinement, additional PO (forward refinement)~\cite{Abrial2010Modeling-in-Eve} are generated to ensure that concrete refinements preserve the abstract models' properties. The Event-B toolset is Rodin~\cite{linkRodin}, result of an EU research project\footnote{RODIN:Rigorous Open Development Environment for Open Systems (EU IST Proj)}: software tool, based on modern software programming tools created to help the development of specifications based on the idea that large complex or critical projects should start with modelling and reasoning about its specification. 

%\subsection{Requirements}

%\subsection{Decomposition}

%\subsection{Generic Instantiation}
%\label{SubSection:GenericInstantiation}

\section{Case study construction}
\label{Section:CaseStudyConstruction}

In this section the steps followed throughout the construction of our model are described. The safety-critical metro system case study describes a formal approach for the development of embedded controllers for a metro\footnote{The Event-B model built is available at \url{http://eprints.ecs.soton.ac.uk/23135/}}. Butler~\cite{RailWaySystem} makes a description of embedded controllers for a railway using classical B. %The railway system is based on the french train system and it was subject of study as part of the european project MATISSE~\cite{MATISSE2003MATISSE:-Method}. 
Our starting point is based on that work but applied to a metro system. That work goes as far as our first decomposition.  We augment it by refining sub-components, adding requirements and instantiating emergency and service doors in carriages.

\subsection{Requirements}

%Requirements are a vital part in the modelling, specification and implementation of a system. 
Requirements analysis~\cite{Sommerville:1997:REG:549198} in systems engineering, encompasses tasks that go into determining the needs or conditions to meet for a new or altered product, taking account possible conflicting requirements of the various stakeholders, such as beneficiaries or users. %It is an early stage in the more general activity of requirements engineering which encompasses all activities concerned with eliciting, analysing, documenting, validating and managing software or system requirements. 
%The definition of the system and its properties 
There are several techniques to deal with requirements and they vary according to projects' domains. Moreover guidelines~\cite{Sommerville:1997:REG:549198} have been developed to achieve this goal. Nevertheless requirements are often described in an informal manner. Consequently it is hard to \emph{reason} about each requirement: experienced people are able to detect contradictions and uncertainties but it is not guaranteed that all will be uncovered. Moreover, within the formal methods domain, it is hard to trace informal requirements with the model/implementation.

Although not available when we developed this case study, a requirement plug-in (ProR~\cite{ProR,HalJasLad2012}) now exists for the Rodin platform, supporting ReqIF 1.0.1 Standard\footnote{ReqIF: Requirements Interchange Format - \url{http://www.omg.org/spec/ReqIF/}}. Benefits of ProR are  incremental creation of hierarchical requirements structures from informal requirements or providing traceability between requirements and formal models. Furthermore, the system description, mixing formal and informal artefacts may contain assumptions about the environment or requirements properties and ProR can reason about them (possibly uncovering contradictions and uncertainties). %Requirements are presented in a table-view and creating annotated links between requirements is supported. 

Our metro system is characterised by trains, tracks circuits (also called sections or CDV and a communication entity (\emph{comms}) that allows the interaction between trains and tracks. The trains circulate in sections and before a train enters or leaves a section, a permission notification must be received. In case of hazard situations, trains receive braking notifications. \emph{Track} is responsible for controlling the sections, changing switch directions (switch is a special section that connects different routes and can be either divergent or convergent) and sending signalling messages to the communication entity.
%
%An overview of the entire development can be seen in Fig.~\ref{Fig:MetroSystemRefDiagram}. After the first decomposition, sub-components can be further refined. Train global properties are introduced in \textit{Train} leading to several refinements until \textit{Train\_M4} is reached. \textit{Train\_M4} is decomposed into \textit{LeaderCarriage} and \textit{Carriage}. We are interested in refining the sub-component corresponding to carriages in order to introduce doors requirements. These requirements are extracted from real requirements for metro carriage doors.\emph{Carriage} is refined and decomposed until it fits in a generic model \textit{GCDoor} corresponding to a \textit{Generic Carriage Door} development as seen in Fig.~\ref{Fig.CarriageRefDiagram}. We then instantiate \textit{GCDoor} into two instances: \emph{EmergencyDoors} and \emph{ServiceDoors} benefiting from the refinements in the pattern. We describe in more detail each of the development steps in the following sections.
%We need to ensure some properties regarding the routes (set of track sections):
These are the main requirements for this case study (some described in Fig.~\ref{Fig:VariablesInvariantsEventsMetroSystemM0}):
\begin{enumerate}
  \item \label{EnumItem:RouteConnected} Route sections are all connected and cannot have empty gaps ($inv1$).
  \item  \label{EnumItem:TransitiveClosure} There are no loops in the route sections: sections cannot introduce loops ($thm3$). 
   %The constant $net$ represents the total possible connectivity of sections (all possible routes subject to the switches positions) defined as relation $CDV \rel CDV$ ($axm1$). 
   Moreover no circularity is allowed (via transitive closure: $thm4$). %: a context is defined and this property is proved over track section relations and functions.).  %as described by $axm2$. Moreover, the no loop property for $net$ is expressed by axiom $axm11$. Theorems $thm1$ states that $net$ preserves transitive closure.
  \item \label{EnumItem:Switches} Switches cannot be connected and can be either divergence or convergent.
  %($axm6$)
  %. They are represented by $aig\_cdv$ divided into two kinds: $div\_aig\_cdv$ for divergence switches and $cnv\_aig\_cdv$  for convergent switches. 
 % They have at most two predecessors and one successor or one predecessor and two successors 
  %($axm10$)
  %.
  \item \label{EnumItem:NonSwitches} Non-switches have at most one successor and at most one predecessor section%($axm9$)
  .
   \item \label{EnumItem:TrainTracks} Trains circulate in tracks ($inv4,inv5, inv7$), %The current route based on current positions of switches is defined by $next$: a partial injection $CDV \pinj CDV$. $next$ is a subset of $net$ ($inv1$) preserving the transitive closure property as described by theorem $thm1$,$thm2$ and does not have loops ($thm3$). 
   %Sections occupied by trains 
   %are represented by variable $occp$. These sections also 
   preserving transitive closure.
    %as seen by $thm4$.
  \item \label{EnumItem:TrainMinSection} Trains occupy at least one section plus a safety distance ($inv4$). %and the section corresponding to the beginning and end of the train is represented by variables $occpA$ and $occpZ$ respectively. Note that $next$ does not indicate the direction that a train is moving in: the direction can be $occpA$ to $occpZ$ or $occpZ$ to $occpA$. These two variables point to the same section if the train only occupies one section ($inv11$).
  \item \label{EnumItem:TrainsAvoidClash} Trains cannot be in the same section at the same time (trains crashed: $inv13$).
   \item \label {EnumItem:MessagesTrainTrack} Comms handles messages exchanged between trains and tracks. Trains heading to an occupied section receive a negative access and braking message.
  %  Messages are exchanged between trains and tracks via the communication entity. Trains moving towards an occupied section receive a negative access message and should brake.
  \item \label {EmergencyBrakeReq} As part of the safety requirements, all trains have an emergency button. 
   \item While the emergency button is enabled, the train cannot speed up (braking).
    \item If a train door is opened, then the train is stopped (in a platform or due to an emergency). In contrast, if the train is moving, then its doors are closed.
%  \item If a train door is opened, that either means that the train is in a platform or there was an emergency and the train had to stop suddenly.
%  \item A train door cannot be allocated to different trains.
  %\item There is a limit to the number of carriages per train.
%  \item Whenever a carriage alarm is activated, then the emergency button of that same train is activated.
%   \item If a train is not in maintenance state, then it must have the correct number of carriages and the leader carriage must be defined already. %Consequently, this is a condition to be verified before the train can change speed.
%  \item If a train is in maintenance state, then it must be stopped.
%  \item If the speed of a train exceeds the maximum speed, the emergency brake must be activated.
%  \item The sum of carriage doors corresponds to the doors of a train. 
\end{enumerate}
%
%Traceability between the requirements and the model is very important. Requirements can be satisfied in the model via axioms  and theorems (for instance, Req.~\ref{EnumItem:TransitiveClosure}), variables and invariants (Req.~\ref{EnumItem:TrainTracks}) or even a collection of events (Req.~\ref{EnumItem:MessagesTrainTrack}) and are spread out across different refinement levels.

\subsection{Abstraction}
Following a ``top-down'' design, the development starts with an \emph{abstraction} model: description that encompasses the main aspects and goals the system intends to answer, obstructing itself from the implementation and other details. Getting a good abstraction is a very hard task requiring an accurate understanding of the system. Moreover the abstraction is the basis of the development playing a crucial role in the entire model. A good abstraction is often not achieved at first attempt even for experienced developers. It  may change throughout the development to fit additional requirements that came into play on a later stage or when, after a few refinements, it does not fit exactly as initially desired. No tools are available that help finding the right abstraction mainly because each system has its specific properties. It often relies on experience and empiric research. Nevertheless we believe that systems can be categorised according to some common properties, architecture and behaviours and therefore having a abstraction template repository could be helpful when starting a model development. Abstraction templates could then be customised according to specific needs.  Unfortunately such repository does not yet exist, requiring further investigation beyond the scope of this paper.

\begin{figure}[tbh]
\centering
%\subfigure[Variables, invariants in \textit{MetroSystem\_M0}]
   {
     \label{Fig:VariablesMetroSystemM0}
     \fbox{
     \includegraphics[scale=0.54]{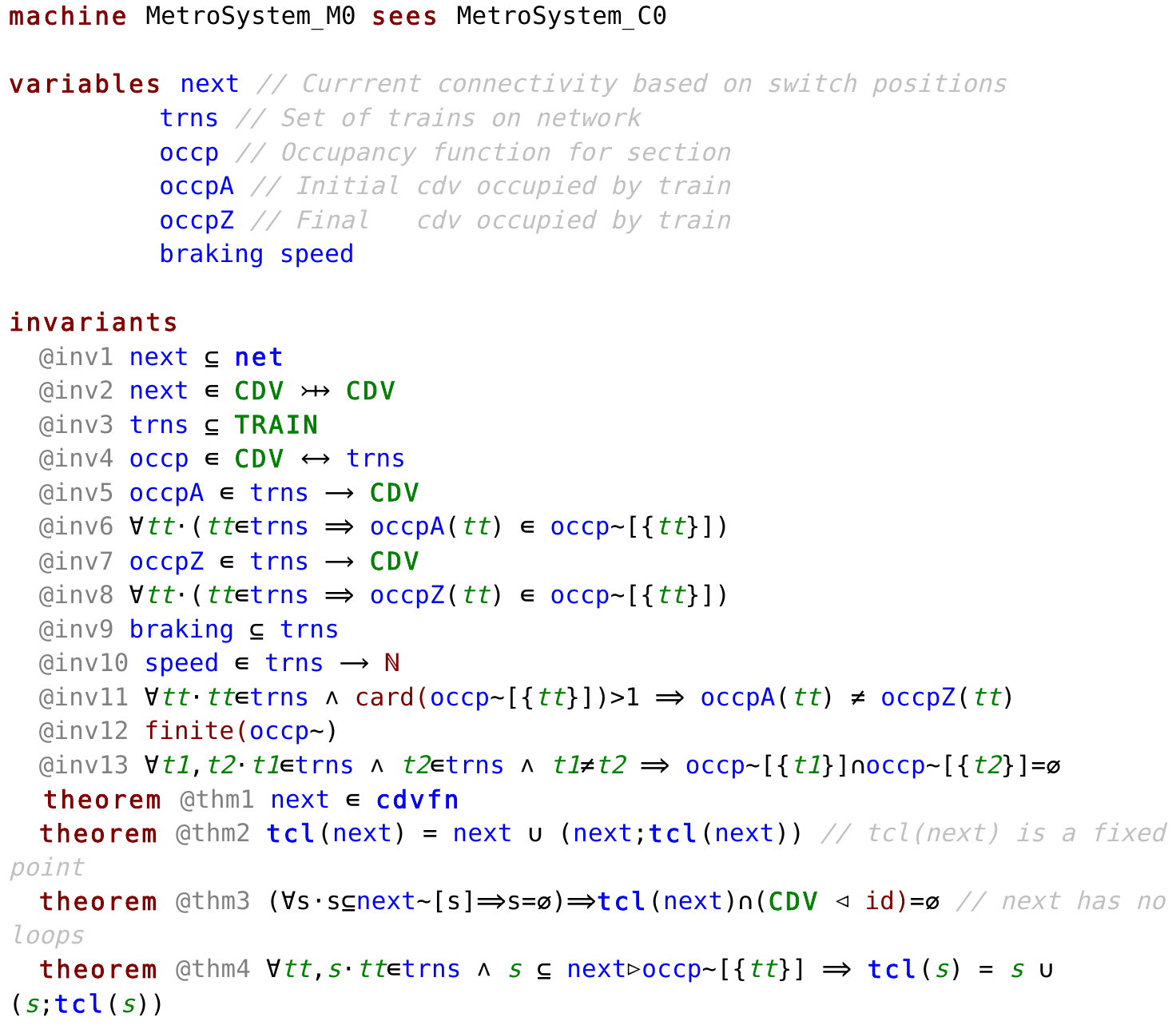}
     %}
     %\fbox{
     %\includegraphics[scale=0.5]{MetroSystem_M0_4}
     }
}   
%\subfigure[Events of \textit{MetroSystem\_M0}]
%    {
%     \label{Fig:EventsMetroSystemM0}
%     \fbox{
%     \begin{tabular}{c}
%     \hspace{-3mm}
%     \parbox{0.394\textwidth}{
%      \vspace{-3mm}
%      \includegraphics[scale=0.55]{MetroSystem_M0_2}
%     }
%     \hspace{12mm}
%     \parbox{0.294\textwidth}{
%      \includegraphics[scale=0.55]{MetroSystem_M0_3}
%     }
%     \hspace{-3mm}
%     \parbox{0.314\textwidth}{
%      \vspace{-32mm}
%      \includegraphics[scale=0.55]{MetroSystem_M0_4}
%     }
%     \end{tabular}
%     }
%}   
\caption{Excerpt of \textit{MetroSystem\_M0}: variables and invariants} 
\label{Fig:VariablesInvariantsEventsMetroSystemM0}
\end{figure}

For our abstraction model (Fig.~\ref{Fig:VariablesInvariantsEventsMetroSystemM0}), we focus on the main properties:  tracks are divided into sections that are connected (Reqs.~\ref{EnumItem:RouteConnected}, \ref{EnumItem:TransitiveClosure}, \ref{EnumItem:Switches}, \ref{EnumItem:NonSwitches}); trains circulate in tracks (Req.~\ref{EnumItem:TrainTracks}); the most important (safety) global property introduced initially states that trains cannot be in the same section at the same time (Reqs.~\ref{EnumItem:TrainMinSection}, \ref{EnumItem:TrainsAvoidClash}). 

\vspace{-10mm}
\subsection{Refinement}
Refinement allows the construction of a model in a gradual way, making it closer to an implementation~\cite{RefDecInst}. At same time, the overall correctness of the system is preserved. 
%A model that maintains the properties of the abstract model and adds more details is defined as a \textit{concrete model}. %abstract model states are linked to the concrete ones. %The refinement process can be repeated so it can be applied over concrete models. 
Our case study heavily uses refinement as seen in Fig.~\ref{Fig:MetroSystemRefDiagram}. At each refinement step, new requirements are introduced to the model and consequently new invariants, variables, events are introduced or refined. 
%
%\begin{figure}[thp]
%\begin{center}
%\includegraphics[scale=0.45]{metroSystemRefinementDiagram}
%\caption{Overall view of the safety-critical metro system development} 
%\label{Fig:MetroSystemRefDiagram}
%\end{center}
%\end{figure}
%
%For refinement \emph{Train\_M1}, carriages are introduced as parts of a train. Each carriage has an individual alarm that when activated, triggers the train alarm (enables the emergency button of the train). Each train has a limited number of carriages. Each carriage has a set of doors and the sum of carriage doors corresponds to the doors of a train. 
%
For instance, for refinement \emph{Train\_M1}, the invariants and properties imposed are:
\begin{enumerate}
  \item There is a limit to the number of carriages per train.
  \item If a carriage alarm is activated, the train's emergency button is also active.
  \item The sum of carriage doors corresponds to the doors of a train. 
  \item Trains have states: maintenance, manual, automatic.
    \item If a train is not in a maintenance state, then it must have the correct number of carriages and the leader carriage must be defined already. %Consequently, this is a condition to be verified before the train can change speed.
  \item If a train is in maintenance, then it must be stopped.
  \item The emergency brake is activated if a train exceeds the maximum speed.
\end{enumerate}
%
%We want to separate the aspects related to carriages from the aspects related to leader carriages:
%%
%\begin{description}
%  \item[Leader Carriage:] Allocates the leader carriage, controls the speed of the train, modifies the state of the train, receives the messages sent from the central, handles the emergency button of the train.
%  \item[Carriage:] Add and removes carriages, opens and closes carriage doors, handles the carriage alarm.  
%\end{description}
%
\subsubsection{Do it right at first/Recursion}
As for abstraction, refinement steps are not reached at first attempt. They evolve, accommodate different requirements and also change, impacting previous refinements. And that comes with a \emph{cost}: a change in the abstraction, affects all the following refinements and the adjustment to each refinement level has to be done manually, which is cumbersome. In our case, the emergency brake requirement (Req. \ref{EmergencyBrakeReq}) was only added after we had reached the first decomposition. The consequences propagated to the abstraction, impacting most events and manual reproving (which delayed for a few days the progress achieved before). This is a limitation of the refinement process in the tool that does not propagate the changes, requiring improvements.

\subsection{Proofs and model construction}
Proofs play an important role in formal modelling, checking that properties and behaviours are preserved. %The complexity/simplicity of the proofs influence how the model is developed. 
There is always a compromise between representing a system, avoiding complex proofs and tool limitations. Despite the plug-ins available for automated proof solving (AtelierB provers~\cite{AtelierB}, Relevance Filter\cite{Roder:2010qy}), complex proofs tend to be avoided. From our experience, a complex proof hard (but not impossible) to discharge, often means that the model is overcomplicated  and may be rewritten/simplified. When building {\small $Train\_M2$}, train doors were represented as {\small $(DOOR\_CARRIAGE;train\_carriage)^{-1}$}, where {\small $DOOR\_CARRIAGE \in DOOR \tfun CARRIAGE$} represents carriage doors and {\small $train\_carriage \in CARRIAGE \\ \pfun trns$} represents the train carriages. Although that relation is enough to describe which doors are part of a train, from a proof viewpoint was very unsuccessful. By rewriting train doors as variable {\small $door\_train\_carriage = (DOOR\_CARRIAGE; \\ train\_carriage)^{-1}$} and invariants {\small $door\_train\_carriage \in trns \rel DOOR, \\ door\_train\_carriage^{-1} \in DOOR \pfun trns$}, we solved the issue. %Afterwards the difficult proofs where easily discharged.

From a tool viewpoint, there is a direct relation between the number of PO per refinement and performance. %The number of properties (usually invariants) to be added for each refinement is relevant. 
Our criteria to choose which properties to add per refinement were directly related with the PO generated per refinement: if over 150 PO, additional properties were stated in new refinements. Train requirements are spread over 4 refinement steps for that reason. Improvements have been made in terms of tool performance in the latest releases but large developments (over 15 refinements and large number of events) are still affected.
%
%\vspace{-15.5mm}
\subsection{Decomposition}
The ``top-down'' style of development used in Event-B allows the introduction of new events and data-refinement of variables during refinement steps. A consequence of this development style is an increasing complexity of the refinement process when dealing with many events and state variables. Model decomposition~\cite{Silva2010Decomposition-T} addresses such complexity by cutting a large model into smaller components. Two methods have been identified for the Event-B decomposition and are supported by a Rodin plug-in~\cite{Silva2010Decomposition-T}: shared variable~\cite{RefDecInst} and shared event~\cite{SyncDecomp}. Because decomposition is monotonic~\cite{SyncDecomp}, the generated sub-components can be further refined independently: sub-components can be used to further refined the original model or be used in other models. Moreover team development can be introduced: different developers can share parts of the same original model by working independently in parallel with the resulting decomposition sub-components.  Decomposition also partition PO which are expected to be easier to discharge in sub-components. In our model, decomposition is used for following reasons: separation of aspects; model architectural decision; tool performance: building/proving is faster for separated models than for monolithics.

\begin{figure}[tbh]
\begin{center}
\includegraphics[scale=0.361]{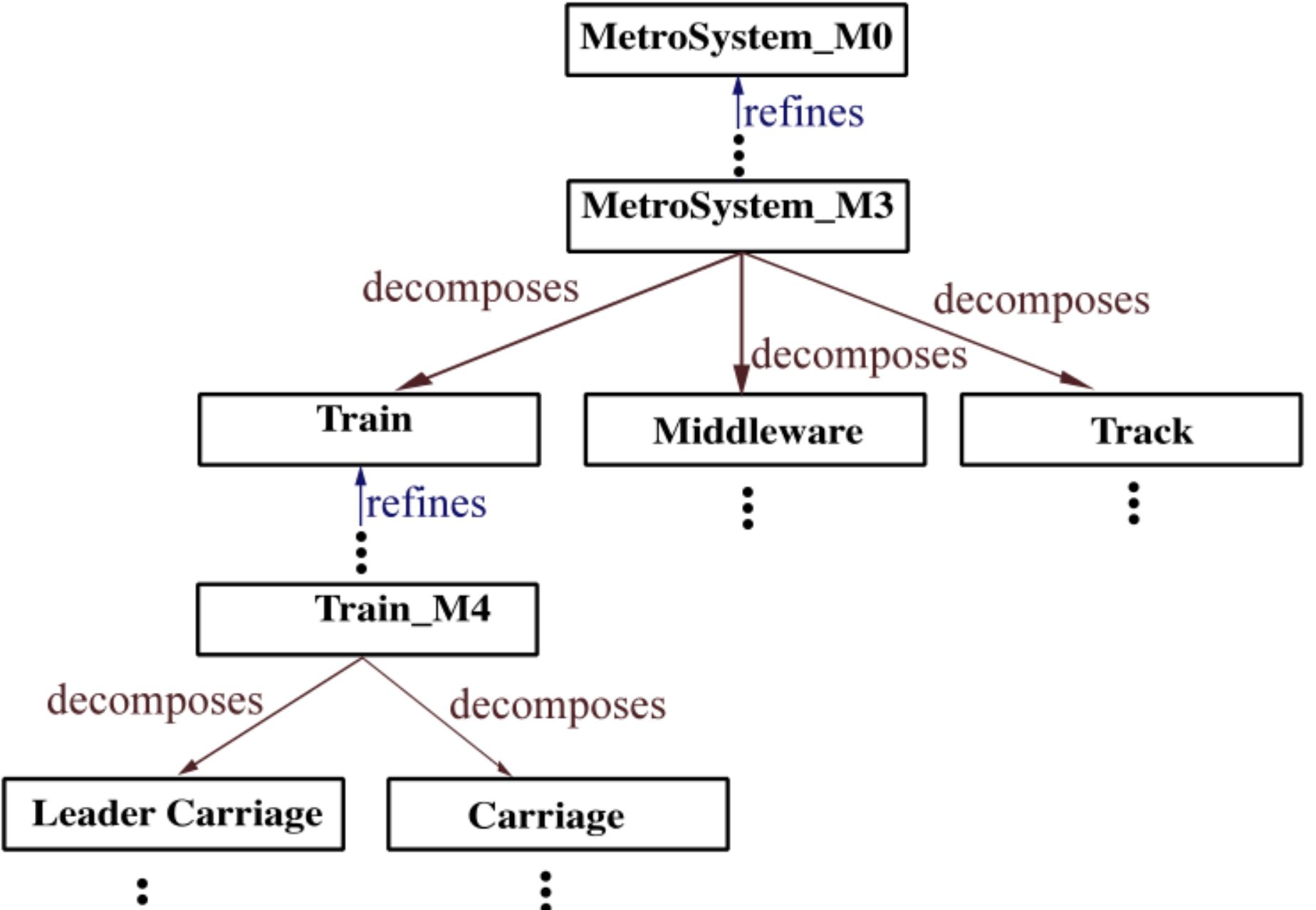}
\caption{Overall view of the metro system development} 
\label{Fig:MetroSystemRefDiagram}
\end{center}
\end{figure}

Decomposition is recursively used as seen in Fig.~\ref{Fig:MetroSystemRefDiagram}: splitting the initial monolithic model into three parts (\emph{Train}, \emph{Middleware} and \emph{Track}) from an architectural point of view (separation of aspects); splitting \emph{Train\_M4} into \emph{LeaderCarriage} (due to the number of POs and separation of aspects) and \emph{Carriage} and later on to decompose \emph{Carriage} into \emph{CarriageInterface} and \emph{CarriageDoor} (Fig.~\ref{Fig.CarriageRefDiagram}). Although we could have used either decomposition styles, we used the shared event style mainly because in that manner, we did not constrain the refinement of variables (like it happens for shared variables). 

Unfortunately the decomposition process does not propagate modifications on the original machine and consequently, decomposed components need to be regenerated if the original component is modified. If the decomposed components have been refined, than the modifications need to be reflected in those refinements (notified via errors or PO being generated or requiring reproving). We believe that the decomposition tool requires improvements in terms of propagation changes to minimise the overall impact that is inevitable.
%
%\vspace{-4mm}
%
%\vspace{-13mm}
\subsection{Generic Instantiation}
\begin{figure}[h]
\centering
\subfigure[Instantiation of $G_{0}..G_{j}$ via parameterisation context $C_{IG}$ creating \emph{instance} $IG$ to fit \emph{problem} $S_{0}..S_{k}$.]
   {
     \label{GenericDevelopment}
    % \fbox{
       \begin{tabular}{c}
       \hspace{-3mm}
     	\includegraphics[scale=0.30]{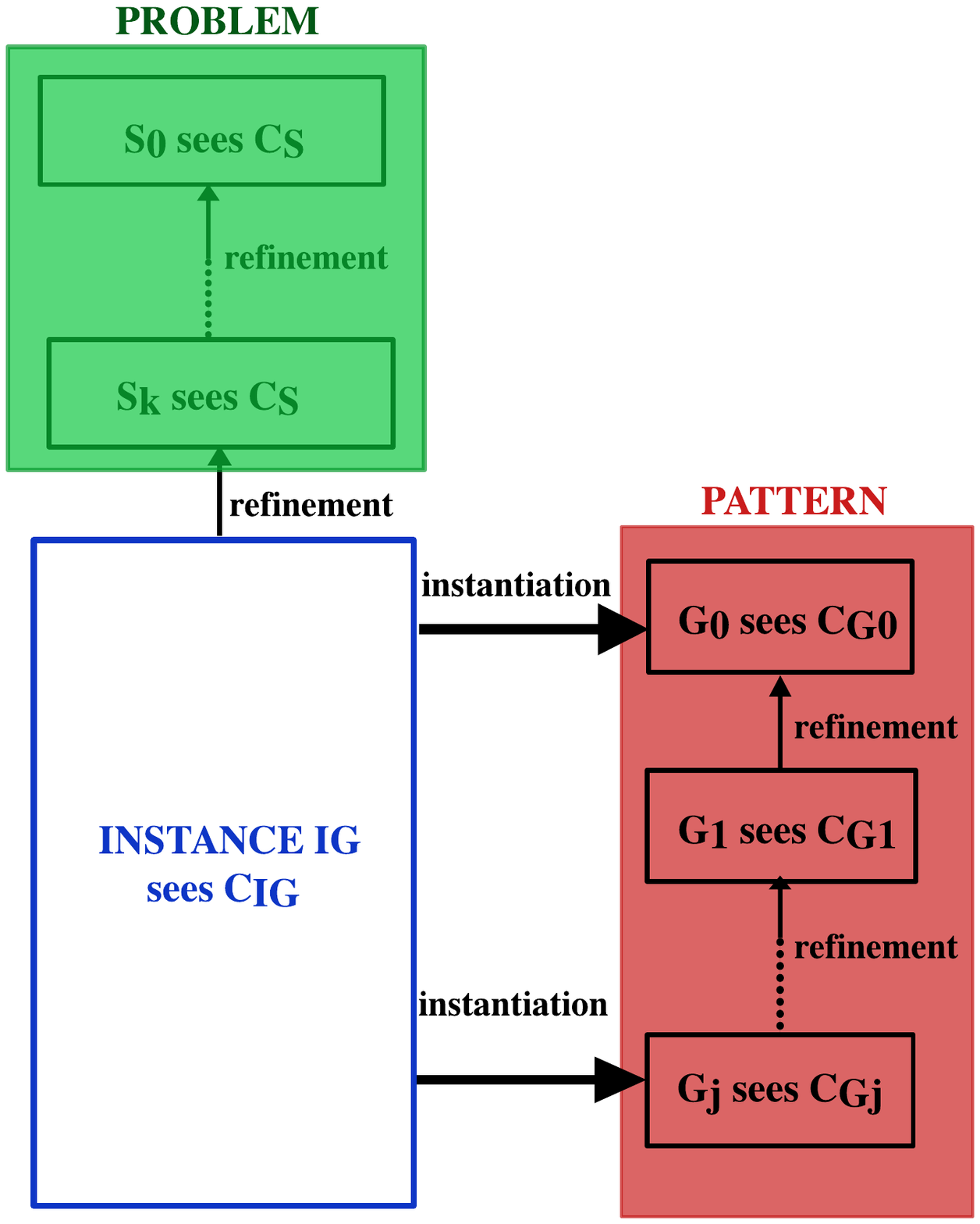}
	\end{tabular}
        %}
    }
\hspace{1mm} 
\subfigure [Carriage Refinement Diagram and Door Instantiation]
   {
     \label{Fig.CarriageRefDiagram} 
 %    \fbox{
     \begin{tabular}{c}
%       \hspace{-1mm}
     	\includegraphics[scale=0.34]{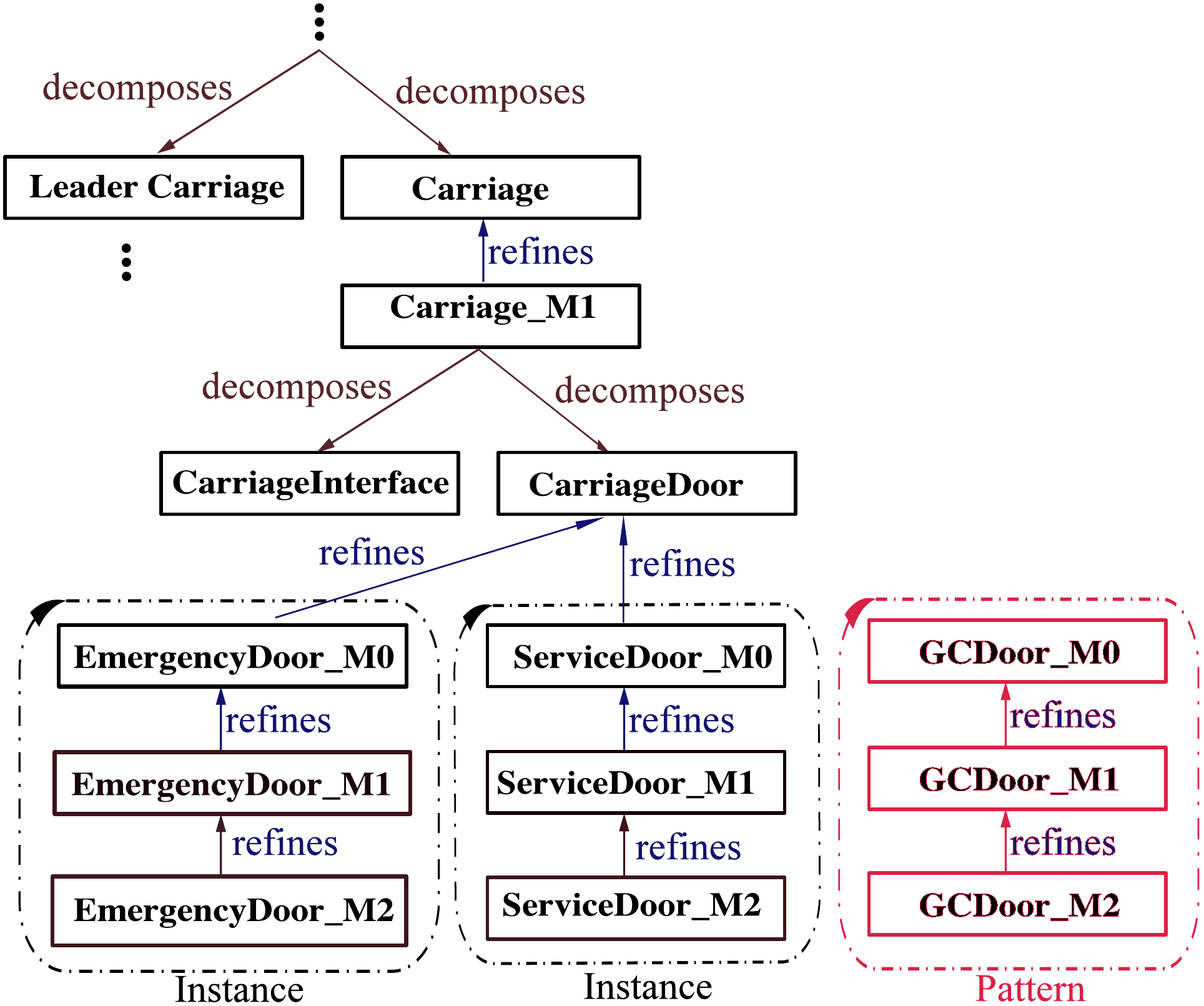}
	\end{tabular}
     %}
    }                      
\caption{Generic Instantiation}
\label{Fig:MachineEmergencyDoorsM2}
\end{figure}

Generic Instantiation can be seen as a way of reusing components and solving difficulties raised by the construction of large and complex models~\cite{RefDecInst}. Generic developments (single machine or a chain of refinements) are reused, originating components with similar properties instead of starting from scratch. Reusability occurs via the instantiation and parameterisation of \textit{patterns}.~\cite{Silva2009SupportingGenericInst} proposes a generic instantiation approach for Event-B by instantiating machines. %The instances inherit properties from the generic development (\emph{pattern}) and afterwards are \textit{parameterised} by renaming/replacing those properties to more specific names according to a particular \emph{problem}. Proofs obligations are generated to ensure that assumptions used in the pattern are satisfied in the instantiation. In that sense the approach avoids re-proving pattern PO in the instantiation. 
The goal is to reuse a \emph{pattern} as an \emph{instance} in an existing development (\emph{problem}) consisting of a chain of refinement of machines $S_{0}$ to $S_{k}$ (\textit{S stands for Specific problem}) as seen in Fig.~\ref{GenericDevelopment}.
%
%\begin{figure}[tbh]
%\centering
%\begin{tabular}{c}
% \includegraphics[scale=0.30]{chain_refinement_generic}
%\end{tabular}
%\caption{Instantiation of \emph{pattern} $G_{0} \dots G_{j}$ via parameterisation context $C_{IG}$ creating instance $IG$ to fit specific \emph{problem} $S_{0} \dots S_{k}$.}
%\label{GenericDevelopment}
%\end{figure}
%
%
\vspace{-7mm}
%\vspace{-5.5mm}
%
The \emph{instance} sees the parameterisation context \textit{$C_{IG}$} (that extends the specific \emph{problem} context $C_{S}$) containing the replacement properties for the elements in context ${C_{Gi}}$. Variables, events and parameters can be renamed to fit new or existing elements in the specific \emph{problem}. %variables $vs$, new events $es$ and new parameters  $ps$. 
%
%The instantiation of refinements reuses the pattern proof obligations in the sense that the instantiation renames and replaces elements in the model but does not change the model itself ( nor the respective properties). 
%
The correctness of the instantiation relies on reusing the \emph{pattern} PO and ensuring that assumptions in the context parameterisation are satisfied in the \emph{instance}.
%
%\subsection{Abstract machine \textit{GCDoor\_M0}}
%
In our case study, an existing development of carriage doors (\textit{GCDoor\_M0}..\textit{GCDoor\_M2}) is used as a \emph{pattern} with all the related PO previously discharged. The \emph{pattern} is instantiated and parameterised accordingly into emergency doors and service doors (Fig.~\ref{Fig.CarriageRefDiagram}).
%
%\begin{figure}[tbh]
%\begin{center}
%\includegraphics[scale=0.34]{CarriageRefinementDiagram}
%\caption{Carriage Refinement Diagram and Door Instantiation} 
%\label{Fig.CarriageRefDiagram}
%\end{center}
%\end{figure}
%
The main \emph{pattern} requirements are:
%
% start by adding the carriage doors and respective states. Four events model carriage doors. The properties to be preserved are:
%
\begin{enumerate}
  \item Doors have a state associated: $open$ (train must be stopped) or $closed$.
  %\item (Carriage) doors can be added or removed from a train.
  \item When adding/removing a carriage to a train, doors must be closed for safety.
  %\item A set of doors can only be closed/opened if the whole set of doors is opened/closed.
  \item Actions involving the doors may result from \emph{commands} ($open$, $close$, $isolate$, $remove\_isolated$) sent from the central door control. %These commands have a state ($start$, $fail$, $success$ and $executed$) and a target (set of doors).
  \item Doors must be closed and locked before a train starts moving.
  \item Doors are opened by the following devices: $manual\_platform$, $manual\_internal$ or $automatic\_central\_door$.
%   If a door is open, then an opening device was used: $MANUAL\_PLATFORM$ if opened manually in a platform, $MANUAL\_INTERNAL$ if opened inside the carriage manually and  $AUTOMATIC\_CENTRAL\_DOOR$ if opened automatically from the central control. 
  \item Doors can get obstructed when closed automatically (people/object obstruction). If an obstruction is detected, a second attempt is made to close them.
   \item Doors can be isolated in case of malfunction or for safety reasons.
 % \item If a command is successful, it means that the command already occurred.
 % \item Two commands cannot have the same door as target except if the command has already been executed.
  \item If a door is obstructed, then it must be in a state corresponding to $open$.
\end{enumerate}

These requirements are shared between both emergency and service doors highlighting the use of instantiation. Additional requirements for each kind of door can be added in further refinements (emergency doors are only available for emergencies, do not respond to standard open command, etc). For our case, the instantiation was manual. Nevertheless currently a generic instantiation prototype is available~\cite{10.1109/SEFM.2009.17}. The tool needs to mature and requires improvements in terms of matching the \emph{pattern} and the last refinement of \emph{problem}. In this case study, the matching was manually achieved through decomposition.
%
%\vspace{-8mm}
\subsubsection{Animation/Model Checker and Code Generation}
Although we are mainly interested in safety properties, ProB model checker~\cite{ProB} proved to be a very useful tool. At some stages, all PO were discharged but ProB showed that the system was deadlocked. In larger developments, these situations may occur frequently. Therefore we suggest safety properties preservation (via PO) and running ProB to confirm deadlock freeness. Another option, to be addressed by ADVANCE\footnote{ADVANCE project: Advanced Design and Verification Environment for Cyber-physical System Engineering- \url{http://www.advance-ict.eu/}} is to introduce liveness properties (e.g. enabledness). Regarding implementation, a code generation plug-in\footnote{Code generation plug-in: \url{http://wiki.event-b.org/index.php/Code_Generation}}~\cite{eps272006} (Event-B to Ada or C) is available.

%Just mention that there are tools available like ProB and Code Generation plug-in for Ada and C! No space for anything else.

\subsubsection{Statistics}
Table~\ref{Tab:StatisticsCaseStudy} describes the statistics of the model in terms of variables, events and PO (including automatically discharged) for each refinement. Almost 3/4 of the PO were discharged automatically.
  \begin{table}[tbh]
  \centering
  \scriptsize
  \begin{tabular}{c}
  $
	\begin{array}{|l|c|c|c|}
      \hline
      & \mathsf{Vars} & \mathsf{Events} & \mathsf{PO/Auto} \\
      \hline
      \textrm{TransitiveClosureCtx} & - & - & 10/10 \\ \hline
      \textrm{MetroSystem\_C0} & - & - & 5/3 \\ \hline 
      \textrm{MetroSystem\_C1} & - & - & 0/0 \\ \hline 
      \textrm{MetroSystem\_M0} & 7 & 10 & 75/64 \\ \hline 
      \textrm{MetroSystem\_M1} & 10 & 13 & 17/17 \\ \hline
      \textrm{MetroSystem\_M2} & 12 & 17 & 78/57 \\ \hline
      \textrm{MetroSystem\_M3} & 12 & 17 & 24/22 \\ \hline
      \textrm{Track} & 4 & 10 & 0/0 \\ \hline
      \textrm{Train} & 7 & 14 & 0/0 \\ \hline
      \textrm{Middleware} & 1 & 4 & 0/0 \\ \hline 
      \textrm{Train\_M1} & 9 & 16 & 74/52 \\ \hline
      \textrm{Train\_M2} & 13 & 21 & 155/79 \\ \hline
          \end{array}
    $
  \end{tabular}
    \begin{tabular}{c}
  $
	\begin{array}{|l|c|c|c|}
      \hline
      & \mathsf{Vars} & \mathsf{Events} & \mathsf{PO/Auto} \\
      \hline
       \textrm{Train\_M3} & 12 & 21 & 65/24 \\ \hline
      \textrm{Train\_M4} & 14 & 21 & 119/89 \\ \hline
      \textrm{LeaderCarriage} & 9 & 21 & 0/0 \\ \hline
      \textrm{Carriage} & 5 & 11 & 0/0 \\ \hline
      \textrm{Carriage\_M1} & 6 & 11 & 28/21 \\ \hline
      \textrm{CarriageInterface} & 4 & 11 & 0/0 \\ \hline
      \textrm{CarriageDoors} & 2 & 5 & 0/0 \\ \hline
      \textrm{CarriageDoorsInst\_M0} & 2 & 5 & 2/1 \\ \hline
      \textrm{GCDoor\_M0} & 2 & 5 & 6/6 \\ \hline
      \textrm{GCDoor\_M1} & 9 & 15 & 81/80 \\ \hline
      \textrm{GCDoor\_M2} & 10 & 22 & 170/153 \\ \hline
      \textrm{Total} &  &  & 909/678 (74.6\%)\\
        \hline
          \end{array}
    $
  \end{tabular}
  \caption{Statistics of the metro system case study}
  \label{Tab:StatisticsCaseStudy}
  %\vspace{-1.9mm}
  \end{table}
 The case study conditions were the following: Rodin v2.1 (Auto Builder: OFF; Auto Prover: OFF), Model Decomposition v1.2.1 and Shared Event Composition plug-in v1.3.1, Generic instantiation was done manually (tool support was not available), ProB v2.1.2.
%\begin{itemize}
%  \item Rodin v2.1 (Auto Builder: OFF; Auto Prover: OFF)
%  \item Model Decomposition v1.2.1 and Shared Event Composition plug-in v1.3.1 
%  \item Generic instantiation was done manually (tool support was not available).
%  \item ProB v2.1.2
%\end{itemize}
%
%\vspace{-5mm}
\section{Related Work and Conclusions}
\label{Section:Conclusions}
From the experience of developing formal models involving a large number of refinements, development tools reach a saturation point where it is not possible to edit the model due to the high amount of resources required (or very slowly). Decomposition is a possible solution that alleviates the issue by splitting the model into tool manageable dimensions, separating concerns, decreasing the number of events and variables per sub-component which results in more manageable models. %Moreover, for each refinement, the properties (added as requirements) are preserved. 
Generic instantiation reuses \emph{pattern} and respective PO per \emph{instance}. 

The experience of modelling a metro system in Event-B using the Rodin platform and its plugins, is shared in terms of model design and assessment of available tools. Requirements are defined and modelled through refinement. As an architectural decision and to alleviate the problem of modelling a  monolithic component, the model is decomposed several times.
%Sub-component \textit{Train} is further refined, introducing several details in four refinements levels. Then again, due to the number of PO, to achieve separation of aspects and to ease the further developments, the model is decomposed into two sub-components: \textit{LeaderCarriage} and \textit{Carriage}.Since we are interested in modelling carriage doors, sub-component \textit{Carriage} is refined. 
Benefiting from an existing development for carriage doors \textit{GCDoor}, this \emph{pattern} is used to instantiate two kind of carriage doors: \emph{service} and \emph{emergency} doors. The refinement of \textit{Carriage} is decomposed, originating \textit{CarriageDoor} that matches with \emph{pattern} \textit{GCDoor\_M0}. Although the instantiation is similar for both cases, the resulting instances can be further refined independently. Generic instantiation minimises the proving effort reusing the pattern \textit{GCDoor} PO (in the overall 257). Therefore we achieve our goal of reusing existing developments and discharging as little PO as possible. Even the interactive proofs were relatively easy to discharge once the correct tactic was discovered. This task would be more difficult without decomposition due to the elevated number of hypotheses to be considered. Nevertheless the effort of discharging PO could be further minimised by having an easy way to reuse PO tactics. A limitation of this model is not addressing liveness properties through proofs which would enrich the model.
% In particular when the same steps are followed to discharge similar POs.

%In a combination of refinement and instantiation, we learned that the abstract machine and the abstract pattern do not necessarily match perfectly. In particular, some extra guards and parameters may exist resulting from previous refinements in the instance. Nevertheless the generic model can still be reused. We can (shared event) compose the pattern with another machine in a way that the resulting events include the additional parameters and guards to guarantee a valid refinement. Another interesting conclusion is that throughout an instantiation, it is possible not to use all the generic events. A subset of generic events can be instantiated in opposition to instantiate all. This a consequence of the event refinements that only depend on abstract and concrete events. Nevertheless this only applies for safety properties. If we are interested in liveness properties, the exclusion of a generic event may result in a system deadlock. 

%With this case study we aim to illustrate the application of decomposition and generic instantiation as techniques to help the development of formal models. Following these techniques, the development is structured in a way that simplifies the model by separating concerns and aspects and decreases the number of PO to be discharged. 
Although we use Event-B, these techniques are generic enough to suit other formal notations and other case studies. Formal methods has been widely used to validate requirements of real systems. 
The systems are formally described and properties are checked to be preserved whenever a system transition occurs. Usually this result in complex models with several properties to be preserved, therefore structuring and reusability are pursued to facilitate the development. Lutz~\cite{Lutz97reuseof} describes the reuse of formal methods when analysing the requirements and designing the software between two spacecrafts' formal models. Stepney \emph{et al.}~\cite{Stepney:2003:OPL:1761968.1761970} propose patterns to be applied to formal methods in system engineering. Using the Z notation, several patterns (and anti-patterns) are identified and catalogued to fit particular kind of models. These patterns introduce structure to the models and aim to aid formal model developers to choose the best approach to model a system, using some examples. Although the patterns are expressed for Z, they are generic enough to be applied to other notations. Comparing with the development of our case study, the instantiation of service and emergency doors corresponds to the Z promotion, where a global system is specified in terms of multiple instances of local states and operations. Although there is not an explicit separation of local and global states in our case study, service and emergency doors states are connected to the state of $CarriageDoor$ and we even use decomposition, instantiation and refactoring %(called meaning preservation refactoring steps in Z promotion) 
to fit into a specific pattern. Stepney~\cite{Stepney:2003:OPL:1761968.1761970} suggests template support and architecture patterns to be supported by tools. We agree and aim to address this issue in the future by having categorised templates customised according to the modeller's needs.

%\subsubsection{Acknowledgments}
%We would like to thank the anonymous reviewers for their constructive comments to improve the quality of the article.

%
% ---- Bibliography ----
 %
\vspace{-1.8mm}
\label{sect:bib}
\bibliographystyle{splncs}

%\bibliography{Mybib}

\begin{thebibliography}{10}

\bibitem{Abrial2010Modeling-in-Eve}
Abrial, J.R.:
\newblock Modeling in Event-B: System and Software Engineering.
\newblock Cambridge University Press (2010)

\bibitem{linkRodin}
Rodin:
\newblock {RODIN project Homepage}.
\newblock \url{http://rodin.cs.ncl.ac.uk} (September 2008) Online; accessed
  27-July-2010.

\bibitem{RefDecInst}
Abrial, J.R., Hallerstede, S.:
\newblock Refinement, {D}ecomposition, and {I}nstantiation of {D}iscrete
  {M}odels: Application to {E}vent-{B}.
\newblock Fundam. Inf. \textbf{77}(1-2) (2007)  1--28

\bibitem{Silva2009SupportingGenericInst}
Silva, R., Butler, M.:
\newblock {Supporting Reuse of Event-B Developments through Generic
  Instantiation}.
\newblock In Breitman, K., Cavalcanti, A., eds.: Formal Methods and Software
  Engineering. Volume 5885 of LNCS.
\newblock Springer Berlin / Heidelberg, Rio de Janeiro, Brazil (December 2009)
  466--484

\bibitem{10.1109/SEFM.2009.17}
Hoang, T.S., Furst, A., Abrial, J.R.:
\newblock {Event-B Patterns and Their Tool Support}.
\newblock 2009 Seventh IEEE International Conference on Software Engineering
  and Formal Methods \textbf{0} (2009)  210--219

\bibitem{Silva2010Decomposition-T}
Silva, R., Pascal, C., Hoang, T.S., Butler, M.:
\newblock {Decomposition Tool for Event-B}.
\newblock Software: Practice and Experience \textbf{41}(2) (February 2011)
  199--208

\bibitem{RailWaySystem}
Butler, M.:
\newblock {A System-based Approach to the Formal Development of Embedded
  Controllers for a Railway}.
\newblock Design Automation for Embedded Systems \textbf{6} (2002)  355--366

\bibitem{Sommerville:1997:REG:549198}
Sommerville, I., Sawyer, P.:
\newblock Requirements Engineering: A Good Practice Guide. 1st edn.
\newblock John Wiley \&amp; Sons, Inc., New York, NY, USA (1997)

\bibitem{ProR}
{ProR}:
\newblock {ProR}.
\newblock \url{http://www.eclipse.org/rmf/pror/} (August 2012) ;Online.

\bibitem{HalJasLad2012}
Hallerstede, S., Jastram, M., Ladenberger, L.:
\newblock {A Method and Tool for Tracing Requirements into Specifications}.
\newblock Submitted to Science of Computer Programming (2012)

\bibitem{AtelierB}
Clearsy:
\newblock {Atelier B Web Page}.
\newblock \url{http://www.atelierb.eu/} (September 2008) Online; accessed
  27-July-2010.

\bibitem{Roder:2010qy}
R{\"o}der, J.:
\newblock {Relevance Filters for Event-B}.
\newblock Masther Theses (2010)

\bibitem{SyncDecomp}
Butler, M.:
\newblock {Synchronisation-Based Decomposition for Event-B}.
\newblock In: RODIN Deliverable D19 Intermediate report on methodology. (2006)
  47--57

\bibitem{ProB}
ProB:
\newblock {ProB}.
\newblock \url{http://www.stups.uni-duesseldorf.de/ProB/overview.php}
  (September 2008) Online; accessed 27-July-2010.

\bibitem{eps272006}
Edmunds, A., Butler, M.:
\newblock {Tasking Event-B: An Extension to Event-B for Generating Concurrent
  Code}.
\newblock In: PLACES 2011. (February 2011)

\bibitem{Lutz97reuseof}
Lutz, R.R.:
\newblock {Reuse of a Formal Model for Requirements Validation}.
\newblock In: In Fourth NASA Langley Formal Methods Workshop. NASA. (1997)

\bibitem{Stepney:2003:OPL:1761968.1761970}
Stepney, S., Polack, F., Toyn, I.:
\newblock {An Outline Pattern Language for Z}.
\newblock In: Proceedings of the 3rd International Conference on Formal
  Specification and Development in Z and B. ZB'03, Berlin, Heidelberg,
  Springer-Verlag (2003)  2--19

\end{thebibliography}

\end{document}